\documentclass[intlimits,twoside,a4paper]{article}

\usepackage[cp1251]{inputenc}
\usepackage[eqsecnum]{cmpj3}
\usepackage{bm}

\issue{2021}{24}{4}{43702}
\doinumber{10.5488/CMP.24.43702}
\title[The half-Heusler compounds]%
{Structural, elastic, electronic and optical properties of the half-Heusler ScPtSb and YPtSb compounds under pressure%
\thanks{Missoum RADJAI: E-mail: mradjai@yahoo.com}}
\author[M. Radjai, A. Bouhemadou, D. Maouche]{M. Radjai\orcid{0000-0002-0313-7155}\refaddr{label1},
        A. Bouhemadou\refaddr{label2}, D. Maouche \refaddr{label2}}
\addresses{
\addr{label1} Laboratory of Physics of Experimental Techniques and Their Applications (LPTEAM), University of Medea, Algeria
\addr{label2} Laboratory for Developing New Materials and their Characterization, University
of Ferhat Abbas Setif 1, Setif 19000, Algeria
}

\date{Received June 20, 2021, in final form August 08, 2021}

\begin{document}

\maketitle

\begin{abstract}%
First-principles calculations using the plane-wave pseudopotential
method within the generalized gradient approximation method were performed
to study the pressure dependence of the structural, elastic, electronic and
optical properties for the half-Heusler compounds ScPtSb and YPtSb
in a cubic MgAgAs-type structure. The calculations were performed with the inclusion of spin-orbit coupling. The calculated equilibrium lattice
parameters are in good agreement with the available experimental and
theoretical values. The crystal rigidity and mechanical stability were
discussed using the elastic constants and related parameters, namely bulk
modulus, shear modulus, Debye temperature, Poisson's coefficient, Young's
modulus and isotropic sound velocities. The calculated electronic band structures show that ScPtSb  has an indirect gap of $\Gamma-X $ type, whereas YPtSb has a direct band gap of $\Gamma -\Gamma $ type. Furthermore, the effect of pressure on
the optical properties, namely the dielectric function, absorption spectrum,
refractive index, extinction coefficient, reflectivity and energy-loss
spectrum is investigated for both compounds ScPtSb and YPtSb.



\end{abstract}%

\section{Introduction}

It is interesting to study ternary half-Heusler compounds due to their specific properties which make them potential candidates for different applications, such as solar cell and thermoelectric
applications~\cite{Graf11}. Considering their crystal structures, the
Heusler alloys are divided into two distinct groups: (i)~full Heusler alloys with
unit formula X$_{2}$YZ and (ii)~the half-Heusler alloys with unit formula XYZ,
where X and Y atoms are transition metals, while Z is either a
semiconductor or a non-magnetic metal~\cite{Graf11}. In recent years, half-Heusler compounds with a narrow band gap and a valence
electron number of $18$ per unit cell have received a great deal of attention due to their
variety and their interesting properties, such  as thermoelectrics, thermal, optoelectronics, and
spintronics~\cite{Gofryk07,Hermet13,Uher99,Hohl99,Larson99,Pierre97,Kandpal06,Larson00,Tobola96,Tobola00}.  Among the numerous reported studies on the half-Heusler compounds, we mention as examples, the thermodynamic properties of TiNiSn compound~\cite{Graf11},
the macroscopic and microscopic properties of LuPtSb compound~\cite{Nowak13}, the elastic, mechanical, phonon and thermodynamic properties of LuAuPb and 
YAuPb compound~\cite{Singh17}, the effect of pressure on the structural,
electronic, elastic, vibration and optical properties of ScXSb 
(X=Ni,~Pd,~Pt) compounds~\cite{Kocak18} and the structural, electronic,
thermodynamic and optical properties of XPtSb (X=Lu,~Sc) compounds~\cite{Narimani16}.

 Despite the large number of studies devoted to the exploration of the properties of the half-Heusler, some fundamental properties of the half-Heusler ScPtSb and YPtSb are insufficiently studied or are not studied at all. The half-Heusler ScPtSb and YPtSb crystallize in the  cubic MgAgAs-type structure in the $F-43m$ (No.~216) space group. ScPtSb and YPtSb compounds obey the $18$ valence electron rule, Sc/Y, Pt and Sb
atoms are positioned at  $4a$, $4c$ and $4b$ Wyckoff atomic positions, respectively. J.
Oestreich and co-workers recently reported the preparation, crystalline structure and experimental band gap of the half-Heusler compound ScPtSb~\cite{Oestreich03}. Knowledge of the fundamental physical properties of a material, such as the elastic, electronic and optical properties, is required to know its eventual application. Ab initio calculations become a very useful tool to predict the fundamental properties of materials. Thus, is the present work, we explored the structural, elastic, electronic and optical properties of ScPtSb and YPtSb compounds through ab initio calculations in the framework of density functional theory. We have also investigated
the pressure effect on the structural, elastic, electronic and optical
properties for ScPtSb and YPtSb. The paper is organized as follows. Calculation details are presented in section 2, the theoretical and computational results
are described in section 3, and finally, the conclusion is presented in the last section.

\section{Computational details}

All first-principles calculations were conducted using the pseudopotential plane-wave (PP-PW) method based on the density functional theory (DFT), as implemented in Cambridge Serial Total Energy Package (CASTEP)~\cite{Clark05}. The exchange-correlation interactions were treated
within the GGA-PBEsol~\cite{Perdew08}. The interactions of valence electrons with ion cores
were represented by Vanderbilt-type ultra-soft pseudopotentials~\cite%
{Vanderbilt92}.The valence-electron configurations for the scandium
(yttrium), platinum, and antimony were generated as Sc(Y):~$3s^{2}$ $3p^{6}$ $%
3d^{1}$ $4s^{2}$ $(4d^{1}$ $5s^{2})$, Pt:~$5d^{9}$ $6s^{1}$ and Sb:~$5s^{2}$ $5p^{3}$, respectively. The plane-wave basis set cut-off energy was set as $650$~eV  and the integration over the Brillion zone was carried out using a $12\times12\times12$ Monkhorst and Pack~\cite{Monkhorst76} k-mesh. The spin-orbit coupling was included. Convergence test demonstrated that these calculation parameters were sufficient to ensure well the convergence of the
total energy, the elastic constants and the density of states (DOS). The optimized structural parameters were determined using the Broyden-Fletcher-Goldfarb-Shanno (BFGS) minimization scheme~\cite{Fischer92} to find the lowest energy structure, with the following
convergence thresholds: energy change per atom less than $5\times10^{-6}$~eV/atom, maximum residual force less than $0.01$~eV/\AA, maximum stress 
within $0.02$~GPa and maximum atom displacement within $5\times10^{-4}$~\AA.

The single-crystal elastic constants $C_{ij}$ were determined through first-principles calculations  of stress-strain data by applying a set of the given homogeneous deformations of finite value and
calculating the resulting stress with respect to the optimization of all the internal
degrees of freedoms~\cite{Clark05}. In order to calculate the optical
properties, the Brillouin zone integration was performed using a dense Monkhorst-Pack grid of $20\times 20\times 20$ k-points. The linear optical coefficients, such as the optical reflectivity, extinction coefficient, refractive index, energy-loss spectrum and absorption coefficient, may be computed    from the values of $\varepsilon \left( \omega \right)$ spectrum.

\section{Results and discussions}

\subsubsection{Structural properties}

\begin{figure}[!h] 
	\begin{center}
       \includegraphics[scale=0.5]{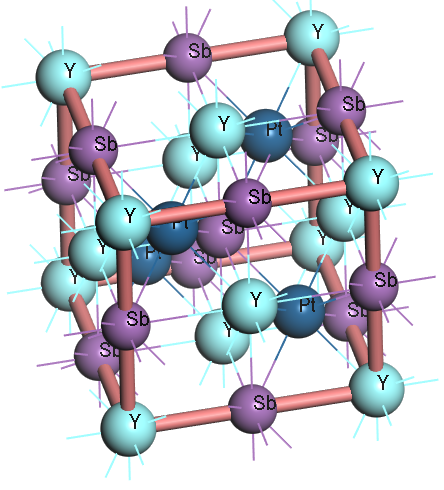}
    \end{center}
	\caption{(Colour online) The unit-cell crystalline structure of the YPtSb compound.}
	\label{fig-1}
\end{figure}

The ternary XPtSb (X=Sc,~Y) half-Heusler compounds crystalize in the cubic MgAgAs-type structure with space group $F-43m$ (No.~216) and $Z=4$. The atomic Wyckoff positions are as follows: X:~$4a$~$(0,0,0)$, Pt:~$4c$~$(1/4, 1/4, 1/4)$ and Sb~$4b$~$(1/2, 1/2, 1/2)$. The crystalline structure of the conventional cell of YPtSb is shown in figure~\ref{fig-1}.
 
\begin{figure}[!h] 
	\begin{center}
       \includegraphics[scale=0.37]{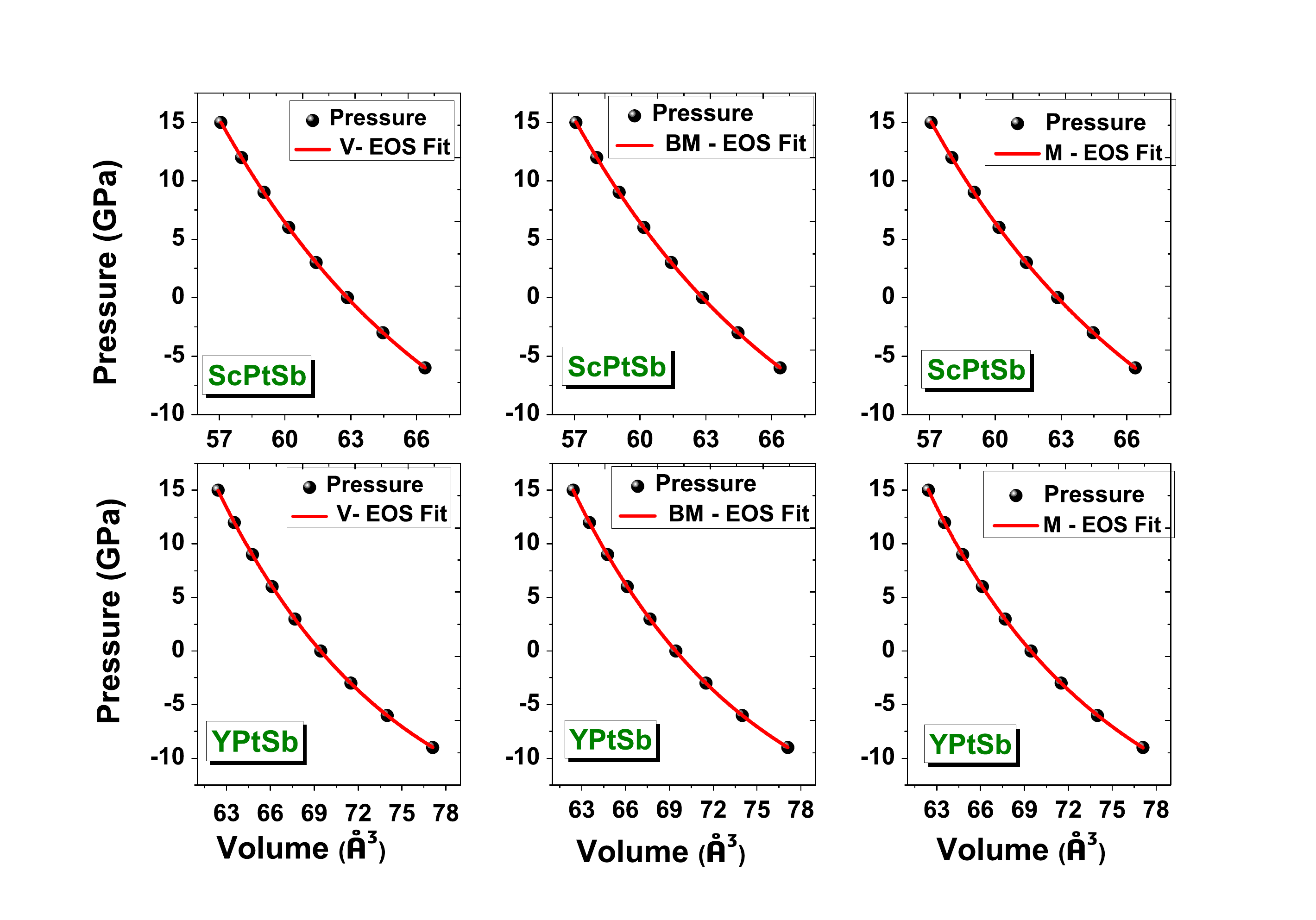}
    \end{center}
	\caption{(Colour online) Calculated pressure vs. volume $P(V)$ for the cubic compouds
ScPtSb and YPtSb. The solid lines are the fits of the obtained data to the
Birch-Murnaghan, Murnaghan and Vinet equations of states.}
\label{fig-2}
\end{figure}

\begin{figure}[!h] 
	\begin{center}
       \includegraphics[scale=0.26]{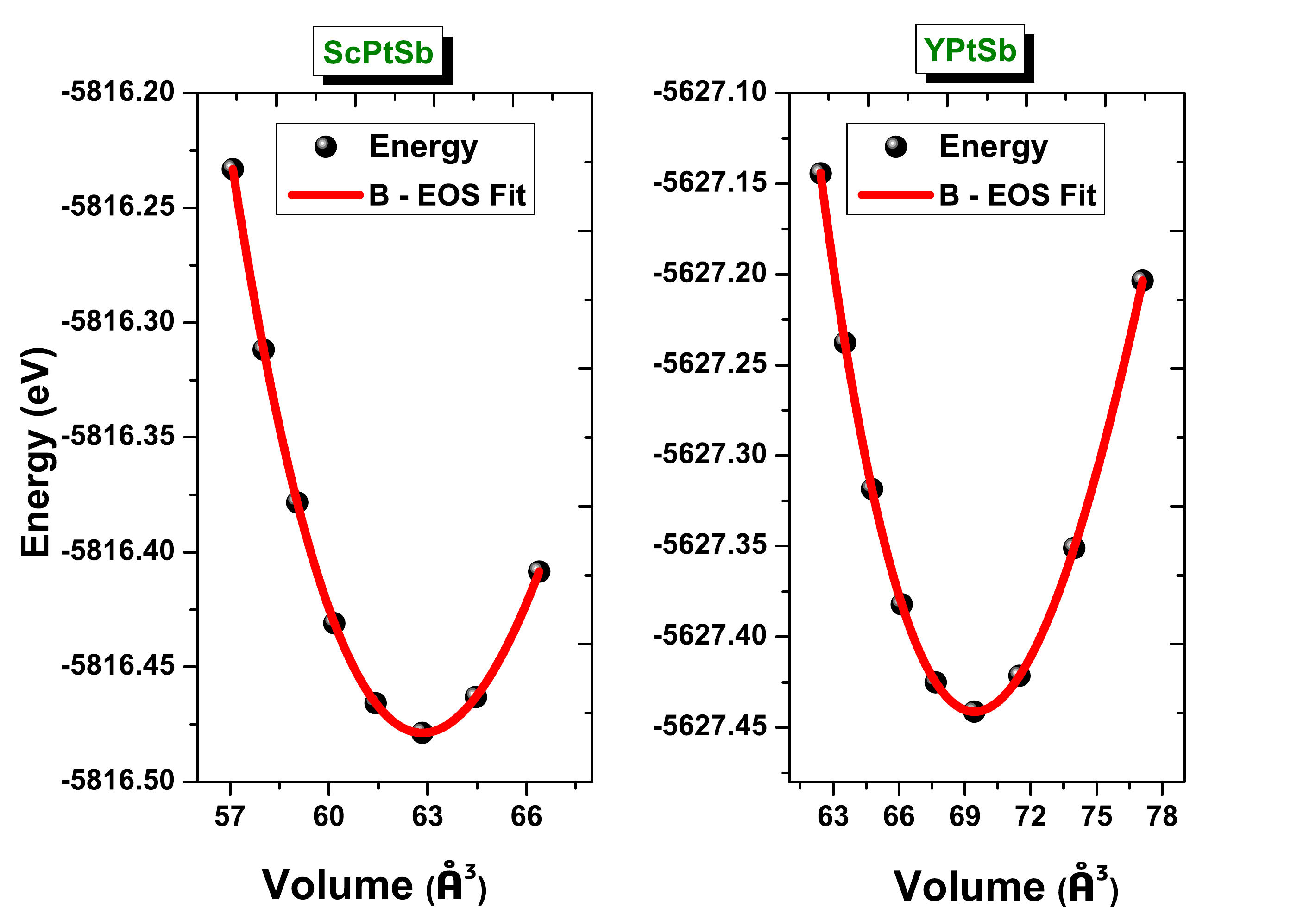}
    \end{center}
	\caption{(Colour online) Calculated total energy vs. volume $E(V)$  for the cubic compouds ScPtSb and YPtSb. The solid lines are the fits of the obtained data to the
Bircht equations of states.}
\label{fig-3}
\end{figure}

The first step of the present work was the calculations of the ground state parameters of ScPtSb and YPtSb,
including the lattice constant $(a)$ and the bond lengths between the constituent's
atoms at $T=0$~K and $P=0$~GPa, using the available experimental results as starting data and the BFGS technique for relaxing the structural parameters. The bulk module $B$ and its pressure derivative $(B^{\prime})$ obtained by fitting the calculated total energy as function of unit cell
volume (E--V) to the Birch equation of state EOS~\cite{Birch47} and by
fitting pressure-volume (P--V) data for different values of the pressure $%
P $ to the Birch-Murnaghan P--V equation of state EOS~\cite%
{Birch78,Ambrosch-Draxl06}, Vinet P--V equation of state EOS~\cite%
{Vinet89,Fu83} and Murnaghan P--V equation of state EOS~\cite{Murnaghan44}%
, given by the following formulas:

\begin{equation}
E\left( V\right) =E_{0}+\frac{9B_{0}V_{0}}{16}\left\{ \left[ \left(
V_{0}/V\right) ^{2/3}-1\right] ^{3}B^\prime +\left[ \left( V_{0}/V\right)
^{2/3}-1\right] ^{2}\left[ 6-4\left( V_{0}/V\right) ^{2/3}\right] \right\} ,
\end{equation}

\begin{equation}
P\left( V\right) =\frac{3}{2}B_{0}\left[ \left( V_{0}/V\right) ^{7/3}-\left(
V_{0}/V\right) ^{5/3}\right] \left\{ 1+\frac{3}{4}\left( B_{0}^{\prime
}-4\right) \left[ \left( V_{0}/V\right) ^{2/3}-1\right] \right\} ,
\end{equation}

\begin{equation}
P\left( V\right) =3B\left( \frac{V_{0}}{V}\right) ^{-{2}/{3}}\left[
1-\left( \frac{V_{0}}{V}\right) ^{-{2}/{3}}\right] \exp \left\{ -\frac{2}{%
3}\left( B^{\prime }-1\right) \left[ \left( \frac{V_{0}}{V}\right) ^{{1}/{3}}-1\right] \right\} ,
\end{equation}%

\begin{equation}
P\left( V\right) =\frac{B}{B^{\prime }}\left[ \left( V_{0}/V\right)
^{B^{\prime }}-1\right] ,
\end{equation}
where $E_{0}$ is the equilibrium energy, $V_{0}$ is the unit-cell volume at
zero pressure. The calculated equilibrium structural parameters are listed in table~\ref{tab:1} along with the available theoretical and experimental  data for the sake of
comparison. The E--V and P--V curves are visualized in figure~\ref{fig-2} and figure~\ref{fig-3}. The calculated lattice parameters for both studied compounds are in very good agreement with the  existing experimental and theoretical data. The calculated lattice parameter of ScPtSb (YPtSb) is somewhat smaller than the corresponding experimental value~\cite{Oestreich03} by about $0.03$\% (0.12\%), while it is somewhat larger than previously calculated value~\cite{Oestreich03} by about $0.58$\% (0.13\%). Our calculated bulk modulus for ScPtSb is somewhat larger than the calculated  value reported  in reference~\cite{Kocak18}, while it is somewhat smaller than that the one reported in~\cite{Narimani16}. One notes that the calculated bulk modulus of YPtSb is somewhat smaller than that of ScPtSb, which indicates that YPtSb is more compressible than ScPtSb and the bulk modulus decreases as the lattice constant increases.

\begin{center}
\begin{table}[!h]%
\caption{Calculated equilibrium crystal lattice constants $a$ (in
\AA),  bulk modulus, $B$ (in GPa), pressure derivative of bulk modulus $B'$
(in GPa)  and bond length (in \AA) for ScPtSb and YPtSb. Compared with the available experimental and theoretical data.}
\label{tab:1}	
\begin{center}
\renewcommand{\tabcolsep}{0.75pc}
\begin{tabular}[b]{|c|c|c|c|}
\hline
\multicolumn{4}{|c|}{\small ScPtSb} \\ \hline
{\small Parameter} & {\small Present work} & {\small Expt} & {\small Other
Cal} \\ \hline
${\small a}$ & {\small 6.310} & 
\begin{tabular}{cc}
{\small 6.312}$^{a}$ & {\small 6.310}$^{b}$%
\end{tabular}
& 
\begin{tabular}{cc}
{\small 6.381}$^{d}$ & {\small 6.273}$^{e}$%
\end{tabular}
\\ \hline
${\small B}_{0}$ & 
\begin{tabular}{cc}
{\small 124.42}$^{1}$ & {\small 124.57}$^{2}$ \\ 
{\small 124.43}$^{3}$ & {\small 124.16}$^{4}$%
\end{tabular}
& {\small -} & 
\begin{tabular}{cc}
{\small 115.137}$^{d}$ & {\small 133.465}$^{e}$%
\end{tabular}
\\ \hline
${\small B}_{0}^{^{\prime }}$ & 
\begin{tabular}{cc}
{\small 4.67}$^{1}$ & {\small 4.76}$^{2}$ \\ 
{\small 4.72}$^{3}$ & {\small 4.63}$^{4}$%
\end{tabular}
& {\small -} & 
\begin{tabular}{cc}
{\small 4.82}$^{d}$ & {\small 5.267}$^{e}$%
\end{tabular}
\\ \hline
{\small Sb-Pt} & {\small 2.732} & {\small -} & {\small 2.733}$^{b}$ \\ 
\hline
{\small Sb-Sc} & {\small 3.155} & {\small -} & {\small 3.156}$^{b}$ \\ 
\hline
{\small Sc-Pt} & {\small 2.732} & {\small -} & {\small 2.733}$^{b}$ \\ 
\hline
\multicolumn{4}{|c|}{\small YPtSb} \\ \hline
${\small a}$ & {\small 6.524} & 
\begin{tabular}{cc}
{\small 6.538}$^{c}$ & {\small 6.532}$^{a}$%
\end{tabular}
& 
\begin{tabular}{cc}
{\small 6.626}$^{f}$ & {\small 6.533}$^{g}$%
\end{tabular}
\\ \hline
${\small B}_{0}$ & 
\begin{tabular}{cc}
{\small 109.73}$^{1}$ & {\small 110.03}$^{2}$ \\ 
{\small 109.70}$^{3}$ & {\small 108.97}$^{4}$%
\end{tabular}
& {\small -} & {\small -} \\ \hline
${\small B}_{0}^{^{\prime }}$ & 
\begin{tabular}{cc}
{\small 4.72}$^{1}$ & {\small 4.69}$^{2}$ \\ 
{\small 4.68}$^{3}$ & {\small 4.67}$^{4}$%
\end{tabular}
& {\small -} & {\small -} \\ \hline
{\small Sb-Pt} & {\small 2.825} & {\small 2.831}$^{h}$ & {\small -} \\ 
\hline
{\small Sb-Y} & {\small 3.262} & {\small 3.269}$^{h}$ & {\small -} \\ 
\hline
{\small Y-Pt} & {\small 2.825} & {\small 2.831}$^{h}$ & {\small -} \\ 
\hline
\multicolumn{4}{|c|}{\small $^{1}$Present work from Birch E-V EOS\cite{Birch47},} \\
\multicolumn{4}{|c|}{\small$^{2}$Present work from Vinet P-V EOS\cite{Fu83},}\\
\multicolumn{4}{|c|}{\small$^{3}$Present work from Birch-Murnaghan P-V EOS\cite{Ambrosch-Draxl06},}\\
\multicolumn{4}{|c|}{\small$^{4}$Present work from Murnaghan P-V EOS\cite{Birch78},}\\
\multicolumn{4}{|c|}{\small$^{a}$Ref~\cite{Oestreich03}, $^{b}$Ref~\cite{Harmening09}, $^{c}$f\cite{Yang08}, $^{d}$Ref~\cite{Kocak18}, $^{e}$Ref~\cite{Narimani16}, $^{f}$Ref~\cite{Li14}, $^{g}$Ref~\cite{Li13}, $^{h}$Ref~\cite{Wenski86}.}\\ 
\hline
\end{tabular}%
	\end{center}
\end{table}%
\end{center}

We calculated the lattice parameter $a$ and the cell volume $V$ under a
series of pressures in order to evaluate the external hydrostatic pressure
effects on the structural properties for both compounds. Variations of $a/a_{0}$ and $V/V_{0}$ as functions of pressure are illustrated in figure~\ref{fig-3},  where $a (V)$ is the lattice parameter (unit cell volume) at the considered pressure and $a_{0}$ ($V_{0}$) is the corresponding value at zero pressure.  The obtained results for $a/a_{0}$ and $%
V/V_{0}$ in the considered range of pressure are well fitted to a
third-order polynomials:

\begin{eqnarray}
\left\{ 
\begin{array}{l}
{a}/{a_{0}}=1-2.65\times10^{-3}P+4.67\times10^{-5}\text{ }P^{2}-6.73\times10^{-7}P^{3}
\\ 
{V}/{V_{0}}=1-7.96\times10^{-3}P+1.60\times10^{-4}P^{2}-2.46\times10^{-6}P^{3},%
\end{array}%
\right. \nonumber\\ 
\left\{ 
\begin{array}{l}
{a}/{a_{0}}=1-3.02\times10^{-3}P+6.06\times10^{-5}\text{ }P^{2}-9.67\times10^{-7}P^{3}
\\ 
{V}/{V_{0}}=1-9.04\times10^{-3}P+2.04\times10^{-4}P^{2}-3.35\times10^{-6}P^{3}.%
\end{array}%
\right.
\end{eqnarray}

The obtained volume compressibility $\beta
_{V}=7.96\times10^{-3}$~GPa$^{-1}$ ($9.04\times10^{-3}$~GPa$^{-1}$) and the linear
compressibilities $\beta _{a}=2.65\times10^{-3}$~GPa$^{-1}$ ($3.02\times10^{-3}$~GPa$^{-1}$) of the lattice parameters $a$, of the both
compounds ScPtSb (YPtSb) respectively, were used to
estimate the bulk modulus $B$ as follows:
\begin{eqnarray}
\left\{ 
\begin{array}{c}
B_{V}={1}/{\beta _{V}} \\ 
B_{a}={1}/{3\beta _{a}}.%
\end{array}%
\right. \nonumber
\end{eqnarray}
The values of the bulk modulus $B$ are $B_{V}=125.62$~GPa (110.61~GPa)
and $B_{a}=125.78$~GPa (110.37~GPa) of the ScPtSb (YPtSb) compounds. These results are in agreement with the bulk module $B$ value
extracted from the equation of state (EOS) fit shown in table~\ref{tab:1}.

\begin{figure}[!h] 
	\begin{center}
       \includegraphics[scale=0.26]{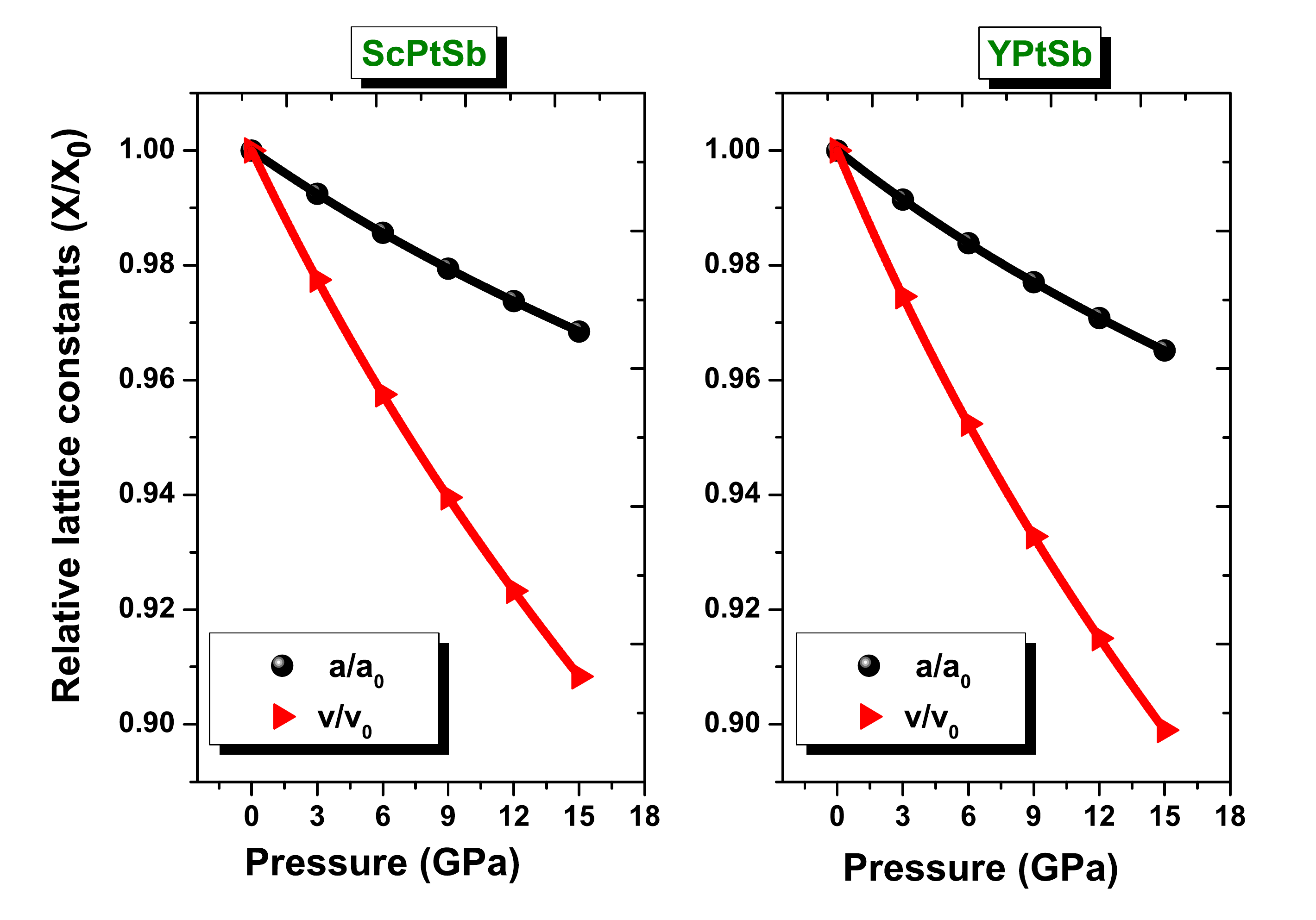}
    \end{center}
	\caption{(Colour online) Pressure dependence of the relative variations of the lattice
constant $a$ and unit cell volume, $V$, for the ScPtSb and YPtSb materials.}
\label{fig-4}
\end{figure}%

\subsection{Elastic properties}

\subsubsection{Single-crystal elastic constants}

\begin{figure}[!h] 
	\begin{center}
       \includegraphics[scale=0.4]{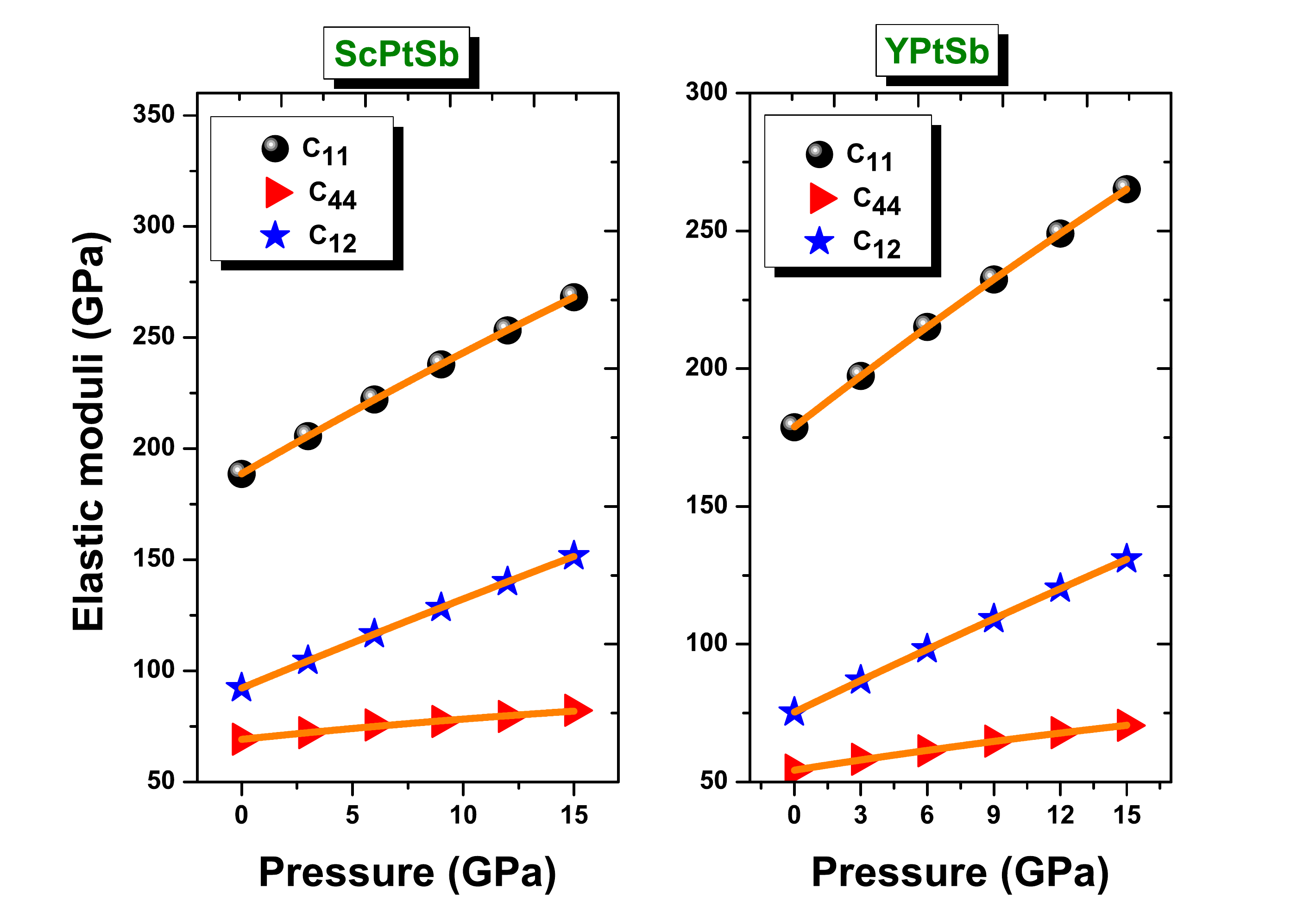}
    \end{center}
	\caption{(Colour online) Calculated pressure dependence of the independent elastic constants
$C_{ijs}$  for the ScPtSb and YPtSb materials. The symbols indicate the calculated results. The lines represent the results of fitting these theoretical results to a second-order polynomial.}
\label{fig-5}
\end{figure}

The elastic properties can be determined from the elastic constants $C_{ij}$. The $C_{ij}$ characterizes the response of a material to an applied stress~\cite%
{Page02} and provides information concerning the nature of the forces
operating in solids~\cite{Westbrook00}. The elastic behaviour of any cubic
crystal is characterized by three independent elastic constants namely, $%
C_{11}$, $C_{12}$ and $C_{44}$. The calculated elastic constants of ScPtSb (YPtSb) under zero pressure and zero temperature are $%
C_{11}=188.62$ ($178.76$)~GPa, $C_{44}=69.05$ ($54.21$)~GPa and $%
C_{12}=92.14$ ($75.26$)~GPa. A cubic crystal is mechanically stable if
its elastic constants satisfy the Born criteria: $C_{11}-C_{12}>0,$ $%
C_{11}>0,$ $C_{44}>0,$ $C_{11}+2C_{12}>0$. The calculated elastic constants of both studied compounds satisfy the above-mentioned criteria, indicating that these compounds are mechanically stable. In order to understand the mechanical stability, phase transition mechanisms, and interatomic
interactions, we studied the effect of pressure on the elastic constants, 
figure~\ref{fig-5} shows the pressure dependence of the three
independent elastic constants of the cubic compounds ScPtSb and YPtSb for
pressures up to $15$~GPa. The mechanical stability criteria under pressure is expressed as follows~\cite{Liu15} $C_{11}-P>\left\vert
C_{12}+P\right\vert ,$ $C_{11}-P>0,$ $C_{44}-P>0,$ $C_{11}+2C_{12}+P>0$.

From the obtained results we can make the following conclusions for both
investigated compounds: (i)~the values of the elastic constants show
that ScPtSb is more resistant to external stress than YPtSb. (ii)~The $C_{11}$, which reflects the resistance to unidirectional compression along the a-axis, is much larger than $C_{44}$, which reflects the resistance to shear deformation, indicating that the studied compounds are more resistive to unidirectional compression than to shear deformation. (iii)~Mechanical stability restrictions are satisfied for both studied compounds in the pressure range $0-15$~GPa. (iv)~Figure~\ref{fig-5} shows that $C_{11}$ is  more sensitive to pressure  than $C_{12}$ and $C_{44}$. The elastic constants $C_{11}$ and $C_{12}$ increase faster with an increasing pressure than $C_{44}$.
Variations of the elastic constants of both studied compounds with pressure were  fitted with the following polynomials of second order for ScPtSb and YPtSb, respectively:

\begin{equation}
\left\{ 
\begin{array}{l}
C_{11}=188.62+5.73P-2.92\times10^{-2}P^{2}, \\ 
C_{44}=69.05+1.04P-1.3\times10^{-2}P^{2}, \\ 
C_{12}=92.14+4.13P-1.17\times10^{-2}P^{2},%
\end{array}%
\right. \quad \left\{ 
\begin{array}{l}
C_{11}=178.76+6.25P-3.37\times10^{-2}P^{2}, \\ 
C_{44}=54.21+1.26P-1.22\times10^{-2}P^{2}, \\ 
C_{12}=75.26+3.86P-1.11\times10^{-2}P^{2}.%
\end{array}%
\right.
\end{equation}

The isotropic elastic moduli of the polycrystalline phase of a material, such as the bulk $B$, shear moduli $G$, Young’s modulus  $E$ and Poisson’s ratio $\nu $, can be calculated from its monocrystalline elastic constants $C_{ij}$ via the well-known Voigt-Reuss-Hill (VRH) approximations~\cite{Voigt28,Hill52}. Voight and Reuss approximations give upper ($B_{V}$ and $G_{V}$) and lower ($B_{R}$ and  $G_{R}$) limits of the $B$ and $G$ moduli, respectively. For a cubic system, the
calculation formulas are as follows~\cite{Wu07}:

\begin{equation}
\left\{ 
\begin{array}{l}
B=\frac{1}{3}\left( C_{11}+2C_{12}\right),  \\ 
G_{V}=\frac{1}{5}\left( C_{11}-C_{12}+3C_{15}\right),  \\ 
G_{R}=[{5\left( C_{11}-C_{12}\right), C_{44}}]/[{3(C_{11}-C_{12})+4C_{44}}],
\\ 
G=({G_{V}+G_{R}})/{2}, \\ 
E={9BG}/({3B+G}), \\ 
\nu =({3B-E})/{6B}.%
\end{array}%
\right. 
\end{equation}

\begin{figure}[!h] 
	\begin{center}
       \includegraphics[scale=0.35]{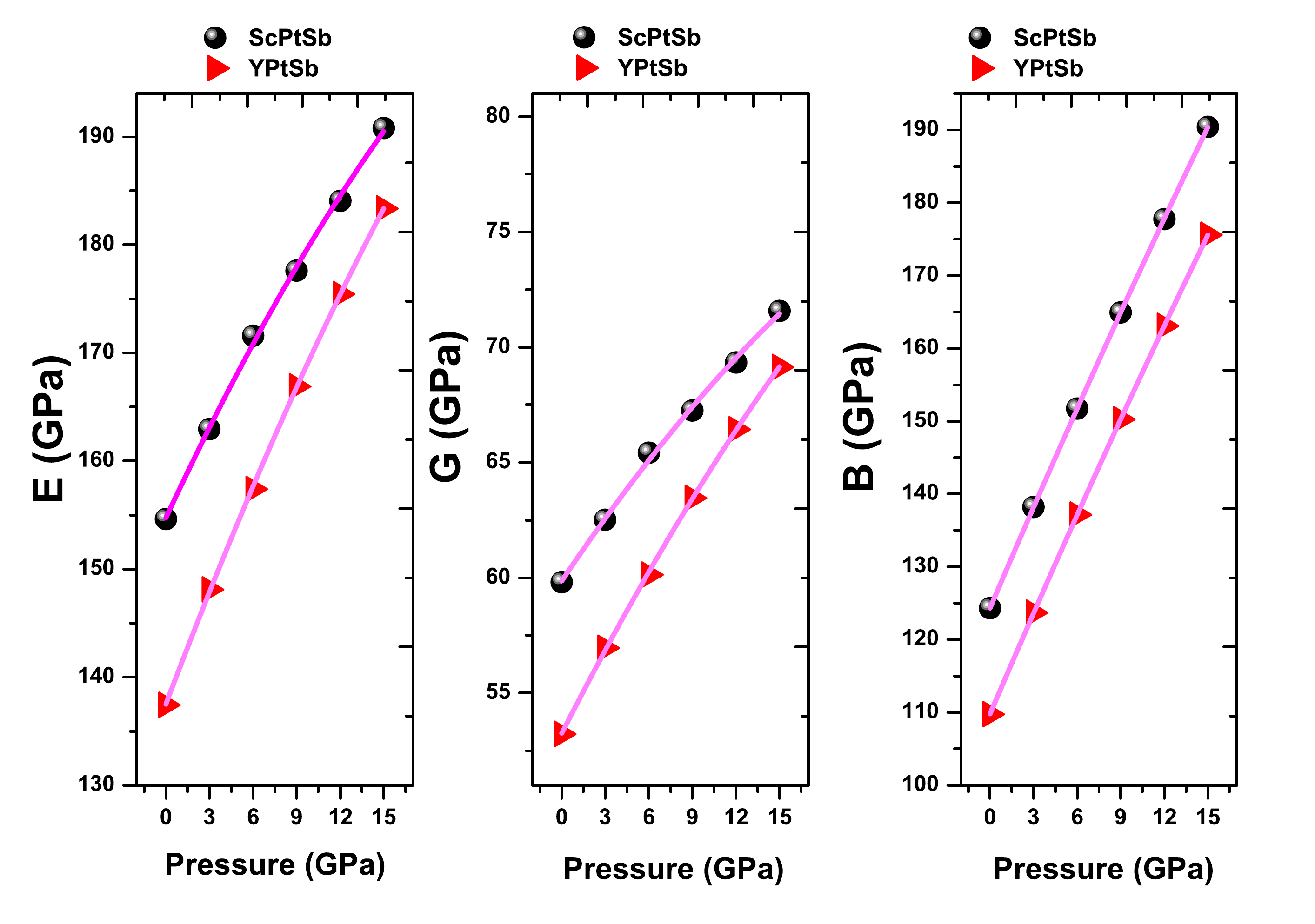}
    \end{center}
	\caption{(Colour online) The values of bulk modulus $B$, shear modulus $G$ and Young's  modulus $E$ under various pressure.}
	\label{fig-6}
\end{figure}

\begin{figure}[!h] 
	\begin{center}
       \includegraphics[scale=0.35]{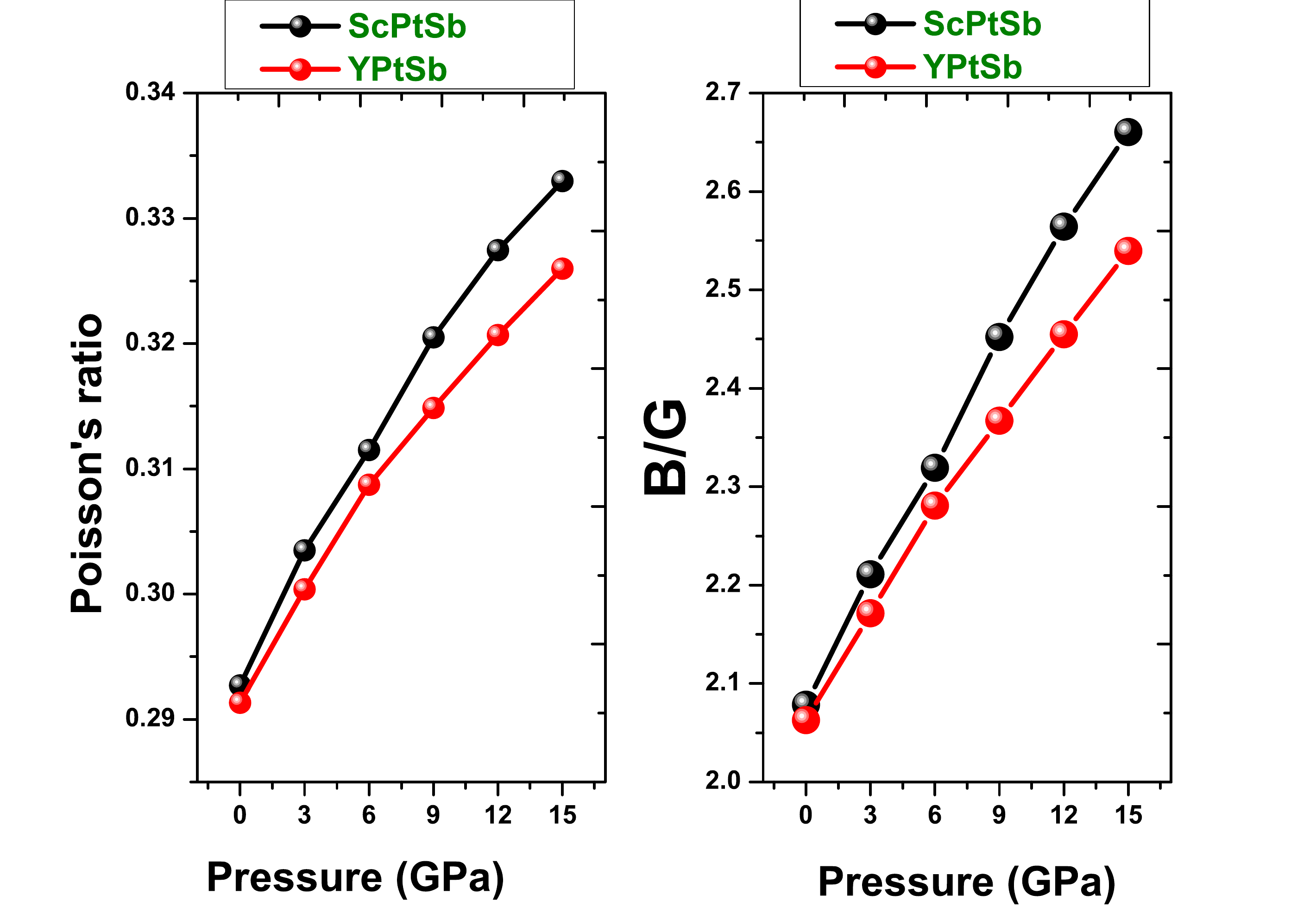}
    \end{center}
	\caption{(Colour online) Calculated pressure dependence of Pugh's ratio $B/G$ and Poisson's ratio $\nu$ for the cubic compounds ScPtSb and YPtSb.}
	\label{fig-7}
\end{figure}

The calculated bulk modulus $B$, shear modulus $G$, Young's modulus $E$ and
Poisson's ratio $\nu $ at zero pressure are quoted in table~\ref{tab:2}. Variations of the
polycrystalline bulk modulus $B$, shear modulus $G$, Young's modulus $E,$ $%
B/G$ ratio and Poisson's ratio $\nu $ as function of pressure are depicted
in figures~\ref{fig-6} and~\ref{fig-7}.

\begin{center}
\begin{table}[!h]%
\caption{Calculated bulk modulus ($B_H$, in GPa); shear modulus ($G_H$, in GPa); Young's modulus ($E$,~in~GPa);
 Poisson's ratio ($\nu$)  and Debye temperature ($\theta_{\rm D}$, in K)  for the ScPtSb and YPtSb
materials.}\renewcommand{\tabcolsep}{1pc}
\label{tab:2}	
\begin{center}
\begin{tabular}[b]{|c|c|c|c|c|c|c|c|}
\hline
System & $B$ & $G_{V}$ & $G_{R}$ & $G$ & $B/G$ & $E$ & $\nu $ \\ 
\hline\hline
ScPtSb & 124.30 & 60.72 & 58.88 & 59.80 & 2.07 & 154.62 & 0.292 \\ \hline
YPtSb & 109.76 & 53.23 & 53.20 & 53.21 & 2.06 & 137.43 & 0.291 \\ \hline
\end{tabular}%
.%
\end{center}
\end{table}%

\end{center}

The obtained results allow us to make the following conclusions:
(i)~the values of bulk modulus~$B$, shear modulus $G$, Young's modulus $E$, $B/G$ ratio and Poisson's ratio $\nu $ increase with an increasing pressure,
indicating that their hardness can be improved at a high  external pressure. (ii)~For both studied compounds, the bulk modulus $B$ is greater than the shear modulus for all the considered pressure range (0--15~GPa). Knowing that the bulk modulus $B$ and shear modulus $G$ mirror the resistance of a material to the volume and shape changes, respectively~\cite{Pugh54}, this result indicates that ScPtSb and YPtSb compounds are more resistant to the volume change than to the shape change. (iii)~Figure~\ref{fig-3} shows that the Young’s modulus $E$ (which characterizes the stiffness of solids) of ScPtSb is larger than that of YPtSb, indicating that ScPtSb is much stiffer than YPtSb. Moreover, the relatively high values of Young’s moduli of both compounds indicate the relatively high resistance to uniaxial deformation (compression/ traction). (iv)~The values of the bulk modulus for ScPtSb and YPtSb compounds deduced from
the single-crystal elastic constants $C_{ij}$ are in good agreement with those calculated from the third order polynomial $P(V)$, and Birch-Murnaghan $P(V$) EOS, Vinet $P(V)$ EOS, Murnaghan $P(V)$, and Birch $E(V)$ EOS (see table~\ref{tab:1}). This is a strong indication of the reliability  of the
calculated elastic constants of ScPtSb and YPtSb and the used computational methodology. (v)~In order to estimate the brittle or ductile behavior of
materials, Pugh~\cite{Pugh54} introduced the criterion of the ratio of the bulk modulus $B$ to
shear modulus $G$. According to this criterion, if $B/G$ value is less than $1.75$ brittle, otherwise the material behaves in a ductile manner. The
calculated $B/G$ values are all higher than $1.75$ for pressures up to $%
15$~GPa as shown in figure~\ref{fig-7}, which suggests that both considered materials are of  ductile nature. The $B/G$ values  of ScPtSb and YPtSb increase with an increasing
external pressure, which suggests that the ductility of these compounds increases with an increasing pressure. (vi)~Poisson's ratio $\nu $ is generally connected with the
volume change in a solid during uniaxial deformation~\cite{Ravindran98}.
Moreover, $\nu $ can provide more information about the characteristics of the
bonding forces~\cite{Nye85}, which usually range from $0.25$ to $0.5$. The calculated values of $\nu $ in the considered pressure range (0--15~GPa) are in the range from $0.26$ to $0.36$, implying that
the interatomic forces are central forces and a considerable volume change can be associated with elastic deformation.

In order to obtain a complete description of the mechanical properties of ScPtSb and YPtSb, we estimated Debye temperature $\theta_{\rm D}$, which is a fundamental parameter closely related to many
physical properties of solids, such as elastic constants, specific heat,
melting temperature, from   the isotropic longitudinal,
transverse and average sound wave velocities ($V_{l}$, $V_{t}$ and $V_{m}$) through the following equations~\cite {Anderson63,Schreiber74,Ravindran98}:

\begin{equation}
\left\{ 
\begin{array}{l}
\theta_{\rm D}=\frac{h}{k_{\rm B}}V_{m}\left( \frac{3n}{4\piup }\frac{N_{\rm A}\rho }{M}%
\right) ^{1/3}, \\ 
V_{m}=\left[ \frac{1}{3}\left( 2V_{t}^{-3}+V_{l}^{-3}\right) \right] ^{-1/3},
\\ 
V_{l}=\left[ (3B+4G)/{3\rho }\right] ^{1/2}, \\ 
V_{t}=\left( {G}/{\rho }\right) ^{1/2}.%
\end{array}%
\right. 
\end{equation}
where, $h$ is Planck constant, $k_{\rm B}$ is Boltzmann constant, $N_{\rm A}$ is
Avogadro number, $\rho $ is the mass density, $M$ is the molecular weight
and $n$ is the number of atoms in the molecule, $B$ is the bulk modulus and $%
G$ is the shear modulus. The calculated values for the sound velocities ($%
V_{l}$, $V_{t}$ and $V_{m}$) and Debye temperature $\theta_{\rm D}$ at zero
pressure are reported in table~\ref{tab3}.

\begin{center}
\begin{table}[!h]%
\caption{Calculated longitudinal, transverse and average sound
velocities ($V_l$, $V_t$ and $V_m$, in m/s) and  mass density  $\rho$ (g/cm$^3$); for the ScPtSb and YPtSb
materials.}\renewcommand{\tabcolsep}{1pc}
\label{tab3}	
\begin{center}
\begin{tabular}[b]{|c|c|c|c|c|c|}
\hline
System & $\rho $ & $V_{l}$ & $V_{t}$ & $V_{m}$ & $\theta_{\rm D}$ \\ 
\hline\hline
ScPtSb & $9.45$ & $4645.31$ & $2514.90$ & $2806.50$ & $301.92$ \\ \hline
YPtSb & $9.24$ & $4422.47$ & $2399.87$ & $2677.67$ & $275.13$ \\ \hline
\end{tabular}%
\end{center}
\end{table}%
\end{center}

\begin{figure}[!h] 
	\begin{center}
       \includegraphics[scale=0.36]{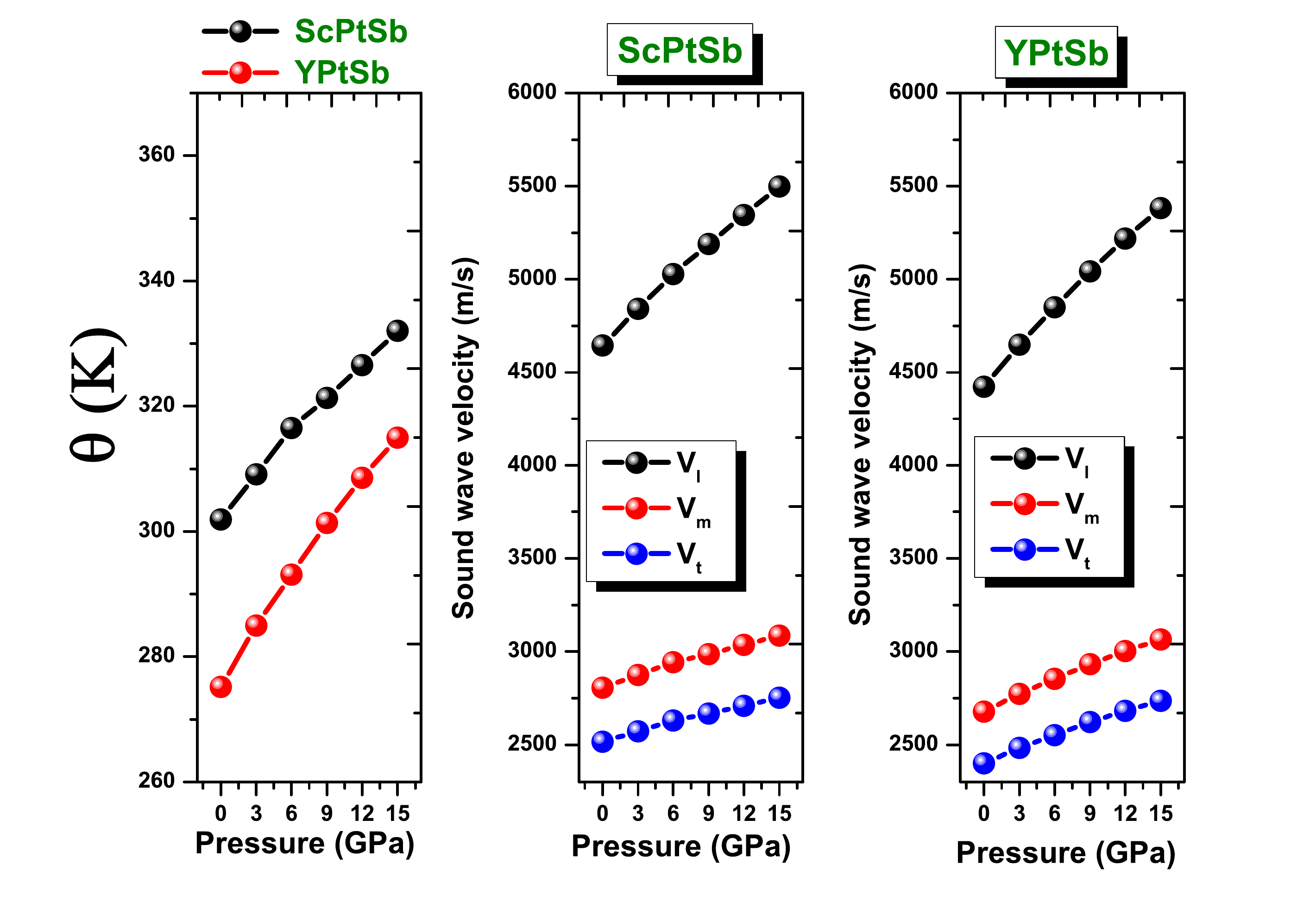}
    \end{center}
	\caption{(Colour online) Calculated pressure dependence of Debye temperature  $\theta_{\rm D}$ and the isotropic sound velocity (longitudinal $V_l$, transverse $V_t$ and average $V_m$) for the cubic compounds ScPtSb and YPtSb.}
	\label{fig-8}
\end{figure}

The calculated values of Debye temperature $\theta_{\rm D}$ and the isotropic sound velocities ($V_{l}$, $V_{t}$ and $V_{m}$) of ScPtSb and YPtSb at different pressures are shown in figure~\ref{fig-8}. The values of the aforementioned parameters increase with an increasing pressure for the ScPtSb and YPtSb compounds. The calculated Debye temperature at zero pressure for ScPtSb is equal to 301.92~K, which is in acceptable agreement with a reported theoretical value of 311.44~K~\cite{Narimani16}. Calculated variation of Debye temperature with pressure is a prediction and there are no corresponding experimental or theoretical results in the literature.  The Debye temperature and the sound
velocity of ScPtSb at $0$~K and $0$~GPa are larger than those of YPtSb. Thus, one can conclude that the propagation velocity sound in ScPtSb is larger than that in YPtSb.
Pressure dependences of $\theta_{\rm D}$ and sound velocity are well
adjusted by a second order polynomial for ScPtSb and YPtSb as follows:

\begin{equation}
\left\{ 
\begin{array}{l}
\theta_{\rm D}=301.92+2.48P-0.034P^{2}, \\ 
V_{l}=4645.31+66.43P-0.65P^{2}, \\ 
V_{t}=2514.90+19.56P-0.27P^{2}, \\ 
V_{m}=2806.50+23.14P-0.31P^{2},%
\end{array}%
\right. \quad \left\{ 
\begin{array}{l}
\theta_{\rm D}=275.13+3.26P-0.041P^{2}, \\ 
V_{l}=4422.47+75.83P-0.79P^{2}, \\ 
V_{t}=2399.87+27.49P-0.34P^{2}, \\ 
V_{m}=2677.67+31.87P-0.39P^{2}.%
\end{array}%
\right.  \label{19}
\end{equation}

It is important to evaluate another parameter, namely the elastic anisotropy
in materials, because it has an important implication in the engineering science as
well as on the nanoscale precursor textures in alloys~\cite%
{Lloveras08,Rong13}, the elastic anisotropy has an important
influence on the induction of microcracks in materials~\cite{Ravindran98,Chung68}. In a
cubic crystal, Zener's $(A_{Z})$~\cite{Zener48} and Every's $(A_{E})$~\cite{Every80} elastic anisotropy factors are defined as follows:

\begin{equation}
\left\{ 
\begin{array}{l}
A_{Z}=2C_{44}/\left( C_{11}-C_{12}\right), \\ 
A_{E}=\left( C_{11}-C_{12}-2C_{44}\right) /\left( C_{11}-C_{44}\right).%
\end{array}%
\right.
\end{equation}

For a completely isotropic material, 
$A_{E}=0$ and $A_{Z}=1$, and thus, the degree of the deviation of $A_{E}$ from zero and $A_{Z}$ from
unity measures the degree of elastic anisotropy. The calculated Zener anisotropy factor is equal to 1.43 for ScPtSb and 1.20 for YPtSb and the Every anisotropy factor is equal to $-0.34$ for ScPtSb and 0.25 for YPtSb, indicating that both studied compounds are characterized by 
a certain degree of elastic anisotropy.

\subsection{Electronic properties}

\begin{figure}[!h] 
	\begin{center}
       \includegraphics[scale=0.35]{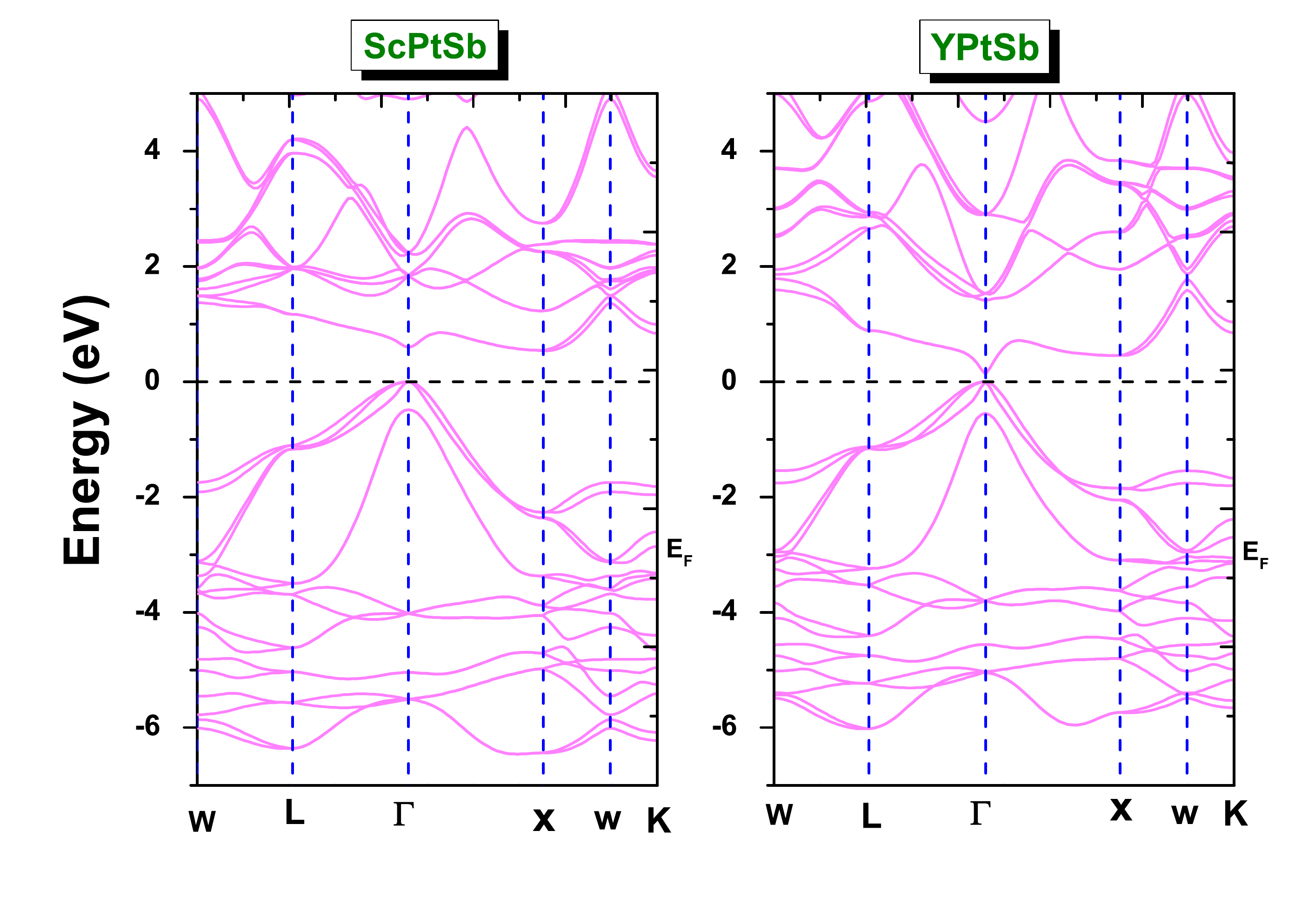}
    \end{center}
	\caption{(Colour online) Electronic band dispersion curves along the high symmetry directions in the Brillouin zone for the ScPtSb and YPtSb materials.}
	\label{fig-9}
\end{figure}

In order to make sense of the electronic behavior of the  ScPtSb and YPtSb half-Heusler compounds, we calculated their band structures along the
high symmetry directions, $\left(W\right.$-$L$-$\Gamma $-$X$-$W$-$\left.K\right)$ in the Brillouin
zone using the GGA-PBEsol approach (see figure~\ref{fig-9}). As can be seen from figure~\ref{fig-9}, the valence band maximum occurs in the $\Gamma $-point for both compounds, while the conduction band minimum occurs at the X-point for ScPtSb and at $\Gamma $-point for YPtSb. Thus, ScPtSb is an indirect band gap semiconductor of $\Gamma -X$ type, while YPtSb is a direct band gap semiconductor of $\Gamma -\Gamma $ type. The calculated fundamental band gap is equal to 0.55~eV for ScPtSb and 0.136~eV for YPtSb. Our calculated band gap for ScPtSb (0.55~eV) is somewhat larger than that of previous calculations (0.49~eV)~\cite{Landau80}. This discrepancy can be attributed to the difference between calculation settings. 


\begin{figure}[!h] 
	\begin{center}
       \includegraphics[scale=0.37]{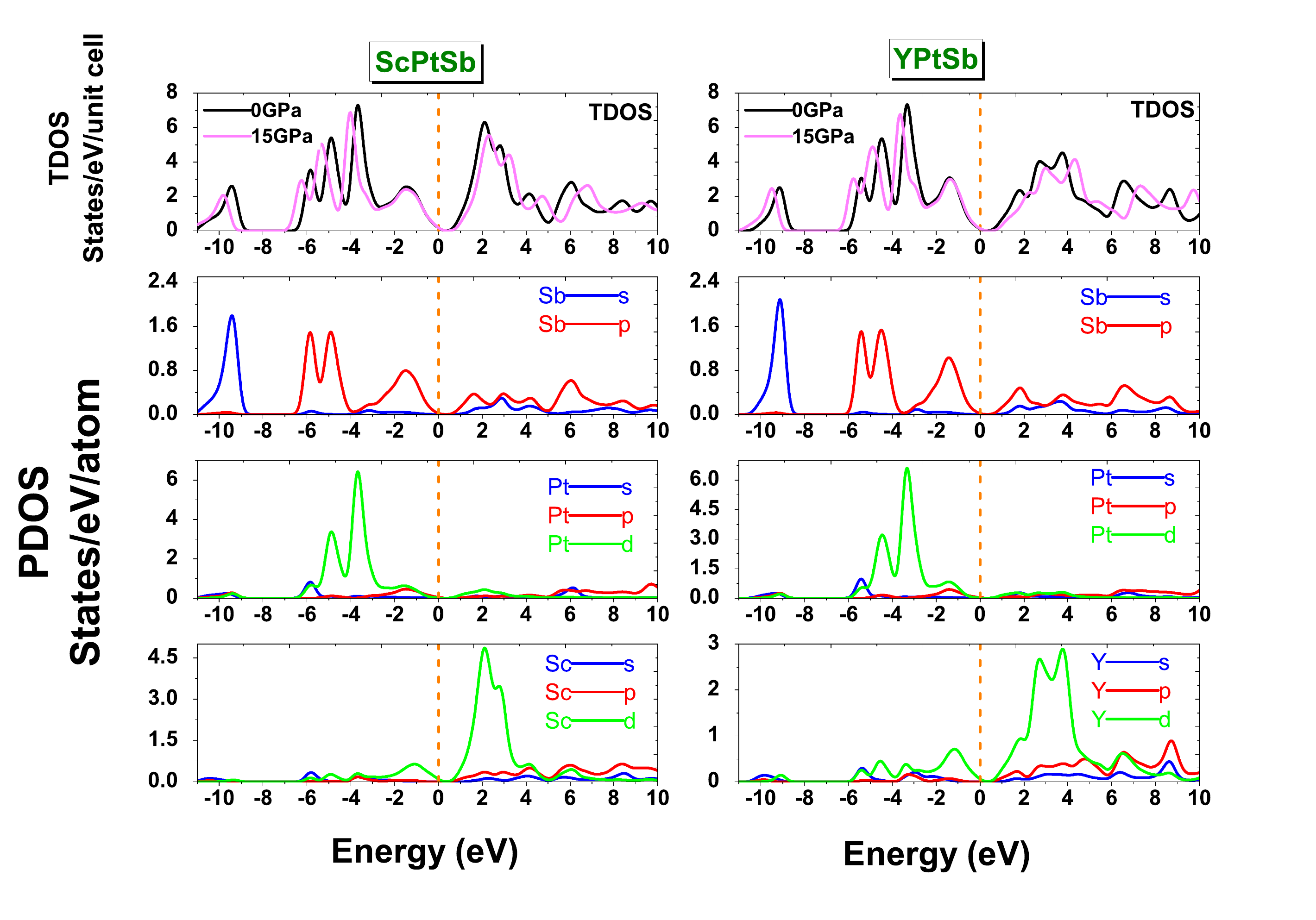}
    \end{center}
	\caption{(Colour online) Total (TDOS) and partial (PDOS) densities of states diagrams for the ScPtSb and YPtSb materials.}
	\label{fig-10}
\end{figure}

Figure~\ref{fig-10} shows the total and partial electron density of states 
(DOS) of ScPtSb and YPtSb  at zero and 15~GPa. The general profiles of both compounds are almost similar with some differences in band splitting and peaks position. The conduction bands of the ScPtSb (YPtSb) compound are dominantly originated from the Sc-$3d$ (Y-$4d$) with small contribution from Sb-$5p$. The  valence bands located between 0 and 12~eV are composed of a mixture of Sc/Y, Pt an Sb electronic states. One notes an hybridization between the Pt-$5d$ and Sb-$5p$, indicating the covalent nature of the Pt--Sb bond. A sharp peak located around $-10.5$~eV and $-10$~eV
originates from the Sb-$5s$ state. As shown in figure~\ref{fig-10} (TDOS), the bonding peaks near the Fermi level gradually reduce the increasing pressure. The band gaps of both studied compounds increase with an increasing pressure. One can see that the shape of TDOS curves presents a slight change at $15$~GPa, which means that half-Heusler compounds do not undergo a structural phase transformation under pressure up to $15$~GPa, and remain structurally stable.

\subsection{Optical properties}

\begin{figure}[!h] 
	\begin{center}
       \includegraphics[scale=0.45]{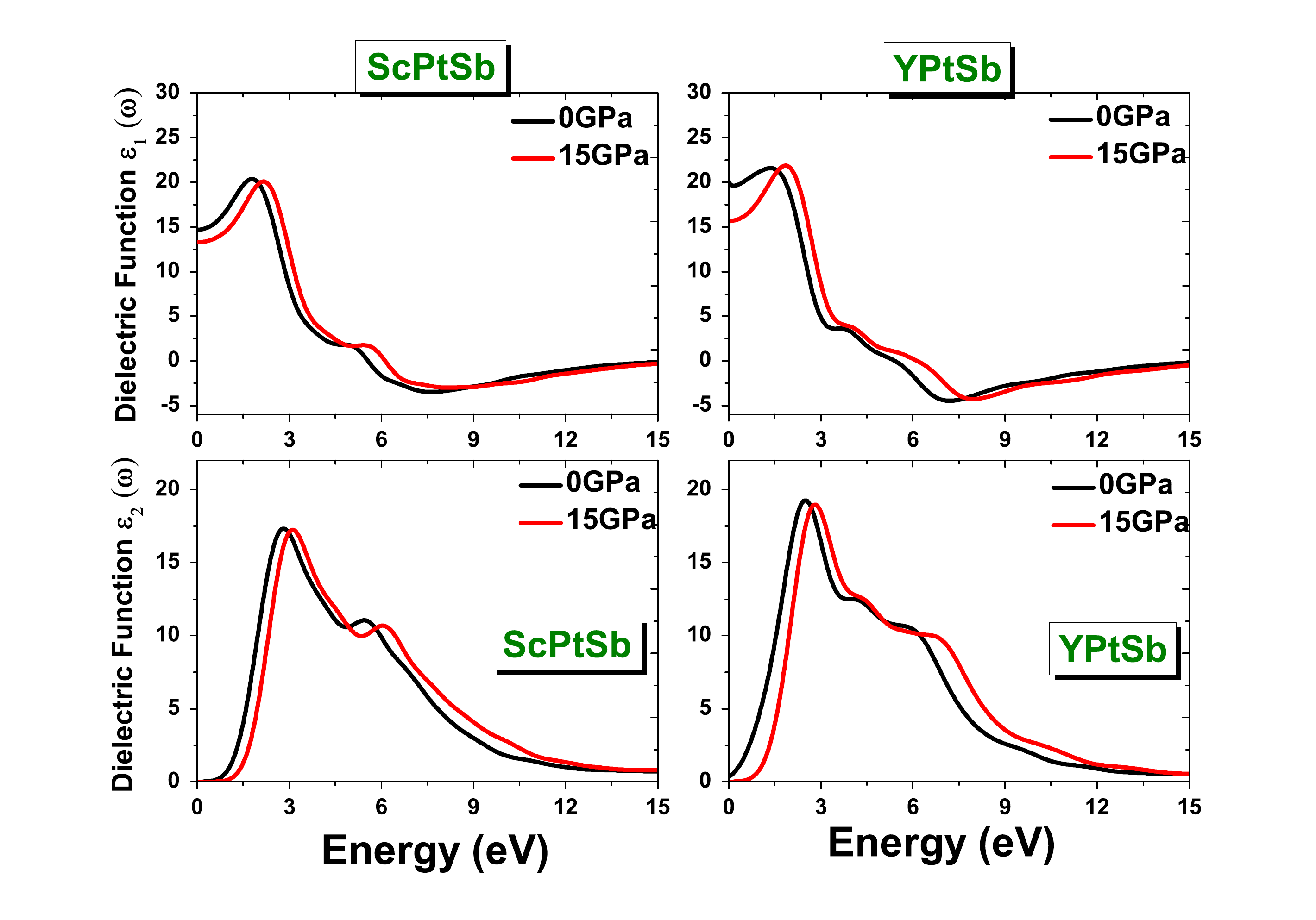}
    \end{center}
	\caption{(Colour online) Calculated imaginary part $\varepsilon_2$ ($\omega$) and real part $\varepsilon_1$ ($\omega$) for the ScPtSb and YPtSb materials.}
	\label{fig-11}
\end{figure}

The optical properties of a solid material are defined by the complex
dielectric function $\varepsilon \left( \omega \right)$ based on the
electronic structure and determined by the transitions between the valence
bands and the conduction bands. In order to understand  the interaction of
photons with electrons~\cite{Sun04} (interaction of radiation  with matter), we calculated the complex dielectric function $\varepsilon \left(
\omega \right) =$ $\varepsilon _{1}\left( \omega \right) +\ri \varepsilon
_{2}\left( \omega \right) $ where the imaginary part of the dielectric
function $\varepsilon _{2}\left( \omega \right) $ describes the optical
absorption in the crystal and $\varepsilon _{1}\left( \omega \right) $ the
real part describes the dispersive part. The real and imaginary parts of
the dielectric function at zero and 15~GPa for ScPtSb and YPtSb are shown in figure~\ref{fig-11} in an energy range from 0 to 15~eV. The zero frequency
limit $\varepsilon \left( 0\right) $ presents an important quantity of $%
\varepsilon \left( \omega \right) $, which depends on the band gap. The calculated value of real part of the dielectric function at the zero-frequency limit $\varepsilon _{1}\left( 0\right)$ is equal to 14.70 (19.88) at zero pressure and 13.14 (15.63) at 15~GPa for ScPtSb (YPtSb). One notes that the $\varepsilon _{1}\left( 0\right) $ value of YPtSb, which has a band gap of 0.13~eV, is lager than that of ScPtSb, which has a band gap of 0.55~eV, indicating that the static dielectric function value $\varepsilon \left( \omega \right) $ is inversely proportional to the band gap value. This trend is consistent with Penn model~\cite{Ahmad15,Penn62}: $\varepsilon _{1}\left( 0\right) =\left( {h\omega _{p}}/{Eg}\right) ^{2}$. The main peak of $\varepsilon _{2}\left( \omega \right) $ spectrum   occurs at $1.78$~eV ($1.23$~eV) at zero pressure and $2.19$~eV
($1.86$~eV)  at $15$~GPa, for ScPtSb (YPtSb). It is noted that
the positions of peaks shift without any important shape change with an increasing pressure
from $0$ to $15$~GPa. It is seen that the general profiles of the $\varepsilon _{2}\left( \omega
\right) $ spectra of the studied compounds are similar.  The highest peak in
the spectrum of the imaginary part of the dielectric function $\varepsilon _{2}\left( \omega
\right) $ (first critical point) for  ScPtSb (YPtSb) is at $2.71$~eV ($2.48$~eV) at zero pressure and $3.12$~eV ($2.84$~eV) at $15$~GPa. The critical point known as the fundamental
absorption edge~\cite{Khenata06}, which gives the threshold for direct
optical interband transition between the topmost valence ($p$ states of
Sb atom) and the lowest conduction band ($d$ states of Sc/Y atoms) is located at around 0.2~eV (0.1~eV) at 0~GPa pressure and 1.06~eV (0.53~eV) at 15~GPa pressure for ScPtSb and YPtSb, respectively.

All-important linear optical constants can be derived from the real and imaginary
parts of the dielectric function $\varepsilon \left( \omega \right) $, namely the
electron energy loss function $L(\omega )$, the refractive index $n(\omega )$, extinction coefficient $k(\omega )$, optical reflectivity $R(\omega )$ and absorption coefficient $\alpha (\omega )$ through the following relations~\cite{Jana13,Wooten72}:

\begin{center}
\begin{equation}
\left\{ 
\begin{array}{l}
n(\omega )=\frac{1}{\sqrt{2}}\left[ \sqrt{\varepsilon _{_{1}}^{2}\left(
\omega \right) +\varepsilon _{_{2}}^{2}\left( \omega \right) }+\varepsilon
_{_{1}}\left( \omega \right) \right] ^{1/2}, \\ 
k(\omega )=\frac{1}{\sqrt{2}}\left[ \sqrt{\varepsilon _{_{1}}^{2}\left(
\omega \right) +\varepsilon _{_{2}}^{2}\left( \omega \right) }-\varepsilon
_{_{1}}\left( \omega \right) \right] ^{1/2}, \\ 
R(\omega )=\left\vert [{\varepsilon \left( \omega \right) ^{1/2}-1}]/[{%
\varepsilon \left( \omega \right) ^{1/2}+1}]\right\vert ^{2}, \\ 
L(\omega )={\varepsilon _{2}\left( \omega \right) }/[{\varepsilon
_{_{1}}^{2}\left( \omega \right) +\varepsilon _{_{2}}^{2}\left( \omega
\right) }], \\ 
\alpha (\omega )=\sqrt{2}\omega \left[ \sqrt{\varepsilon _{_{1}}^{2}\left(
\omega \right) +\varepsilon _{_{2}}^{2}\left( \omega \right) }-\varepsilon
_{_{1}}\left( \omega \right) \right] ^{1/2}.%
\end{array}%
\right. 
\end{equation}
\end{center}

\begin{figure}[!h] 
	\begin{center}
       \includegraphics[scale=0.3]{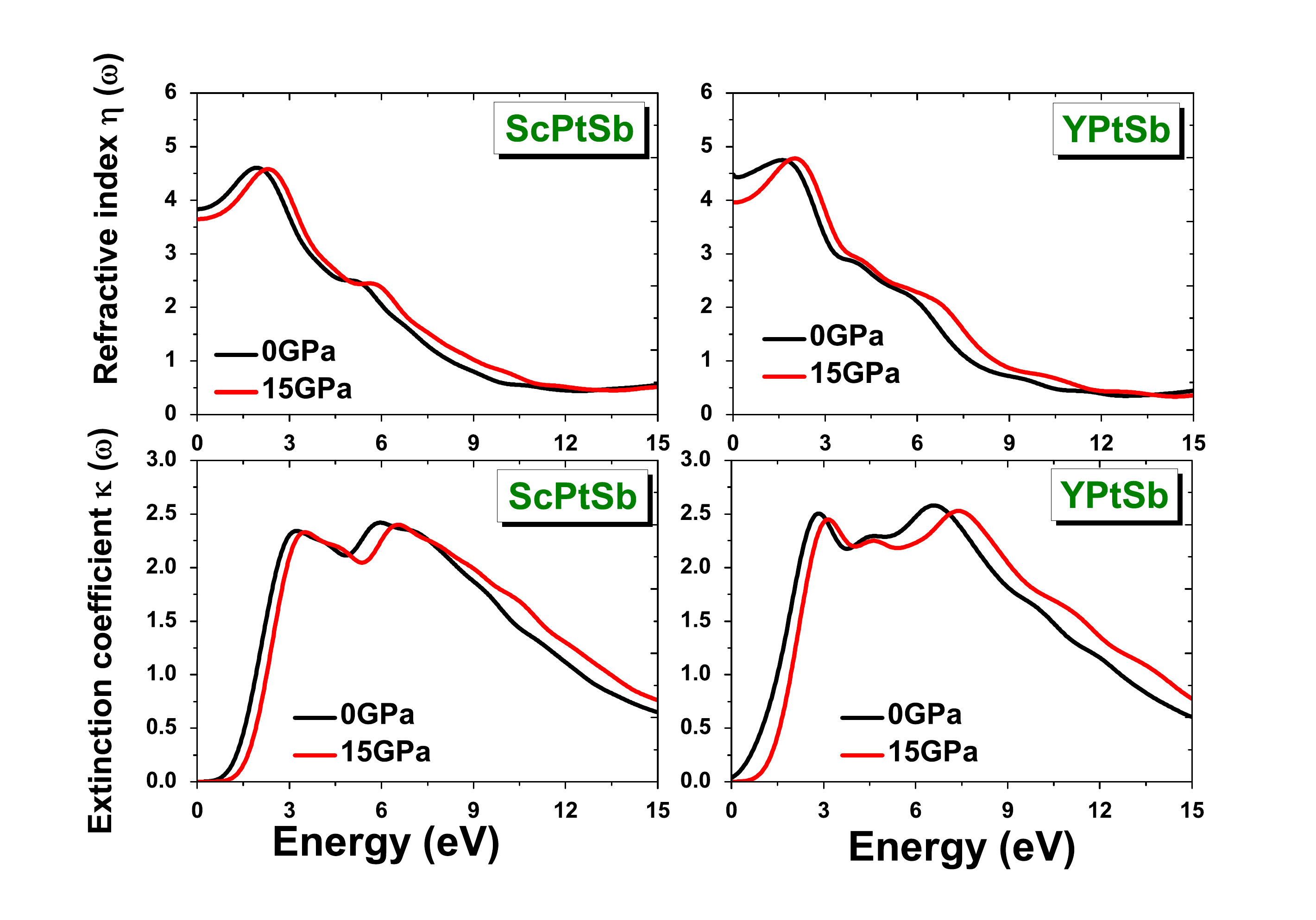}
    \end{center}
	\caption{(Colour online) Calculated refractive index  $\eta$ ($\omega$) and extinction
coefficient  $\kappa$ ($\omega$) for the ScPtSb and YPtSb materials.}
\label{fig-12}
\end{figure}

\begin{figure}[!h] 
	\begin{center}
       \includegraphics[scale=0.3]{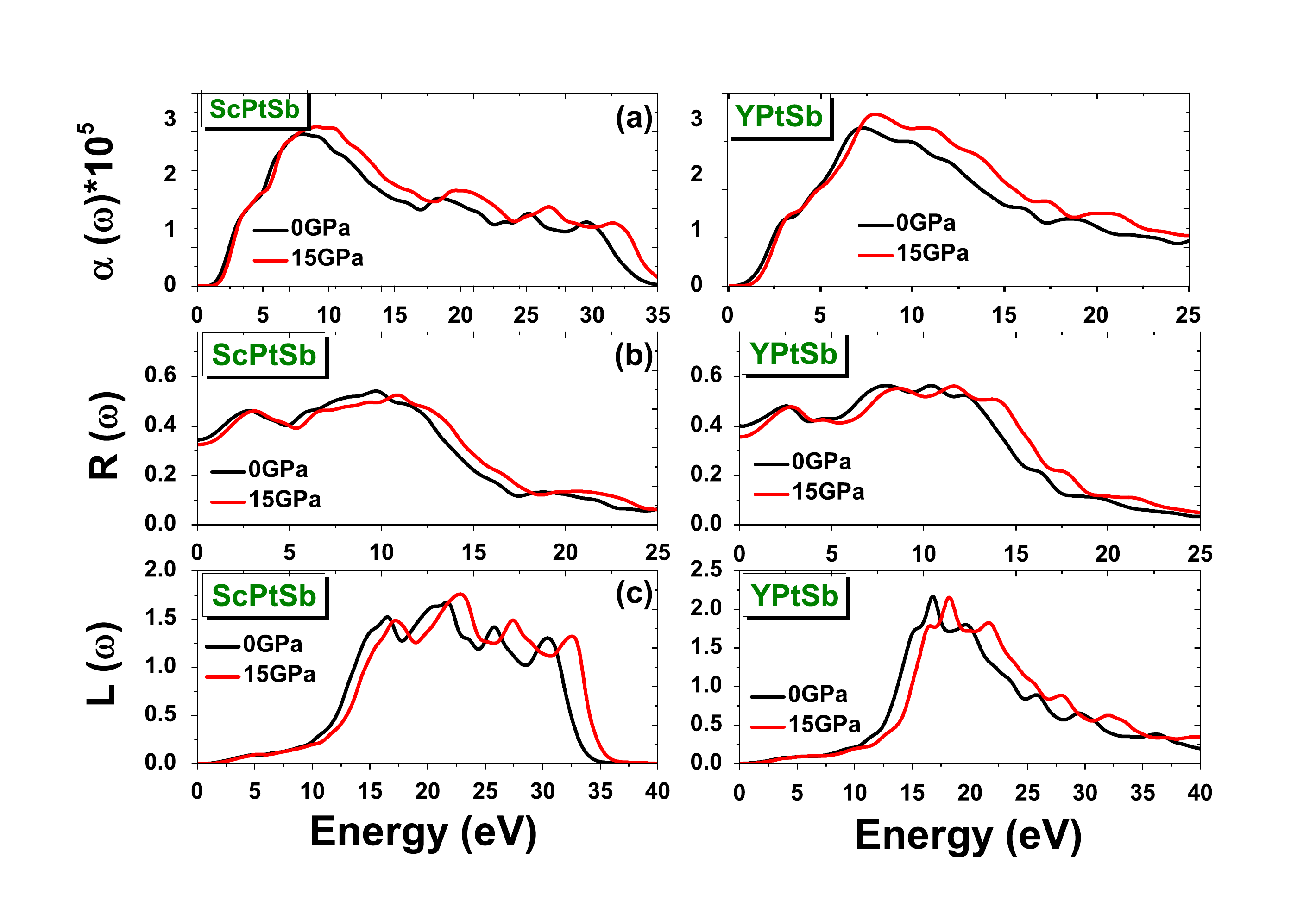}
    \end{center}
	\caption{(Colour online) Calculated optical constants for the ScPtSb and YPtSb
materials: (a) absorption, (b) reflectivity and (c) energy-loss spectrum.}
\label{fig-13}
\end{figure}

Figures~\ref{fig-12} and~\ref{fig-13} show the calculated spectra of the absorption coefficient, refractive index, extinction coefficient, reflectivity and loss-energy function in an energy range from 0 to 15~eV at zero pressure and 15~GPa. To the best of our knowledge, there are no available theoretical or experimental optical spectra in the literature for YPtSb.  

As can be seen from figure~\ref{fig-12}, the value of refractive index at zero pressure $n(0)$ for ScPtSb (YPtSb) is equal to $3.84$ ($4.43$) at zero pressure and $3.60$ ($3.93$) at $15$~GPa, indicating that the static refractive index $n(0)$ decreases with an increasing pressure. The refractive index $n(\omega )$ decreases to attain a minimum level of $0.33$ ($0.26$) at $11.74$ ($12.37$)~eV at zero pressure and at $12.62$ ($15$)~eV at 15~GPa for ScPtSb (YPtSb). The extinction coefficient $k(\omega )$ reaches a maximum of  $2.31$ ($2.50$) at approximately 3.12~eV (2.81~eV) at zero pressure and 2.32 (2.43) at approximately 3.48~eV ($3.11$~eV) at 15~GPa pressure for ScPtSb (YPtSb).

The absorption coefficient is an important optical parameter which provides useful information on the relative absorption of the radiation energy per length unit inside a medium~\cite{Roknuzzaman13}. Figure~\ref{fig-13}~(a) shows that the absorption spectrum
starts to increase when the photon energy is higher than the absorption edge, which is the typical characteristic of semiconductors and insulators compounds. The absorption spectrum increases rapidly with the wide
absorption spectrum of ScPtSb (YPtSb) compounds located about 0--35~eV at
zero pressure and about 0--36~eV at $15$~GPa pressure, showing the interband
transitions from occupied states in valence bands to unoccupied states in
conduction bands. These compounds are not a transparent crystal in the
interval 0--35~eV but present a noticeable absorption in the visible and
far-ultraviolet range, so where the reflectivity and absorption of these
compounds become very small (above $36$~eV), these compounds become
transparent. The maximum absorption of ScPtSb (YPtSb) occurs in the
interval of 7.22--8.88~eV (6.46--8.01~eV) at zero pressure and 7.78--10.95~eV 
(7.2--8.96~eV) at $15$~GPa pressure. The reflectivity spectrum depends
mainly on the incident photon energy. The calculated reflectivity of this $R(\omega)$ spectrum of ScPtSb (YPtSb) is shown in figure~\ref{fig-13}~(b). The static reflectivity at zero frequency of ScPtSb compound is less than the corresponding value of YPtSb, the maximum reflectivity value of about $0.54$ ($0.56$) for photon energies $9.71$~eV ($7.86$~eV) at zero pressure and
about $0.52~(0.56$) for photon energies $11.08$~eV ($11.56$~eV) at $15$~GPa
pressure for the both compounds ScPtSb (YPtSb), respectivily. The
reflectivity spectra decrease abruptly of these compounds after these energy
ranges and reach zero above $35$~eV, due to the excitations of collective
plasma resonance. To further study, we have investigated the $L(\omega )$
loss energy function, is a significant fundamental factor to explain the
energy loss of a fast electron traversing in a material, such as
interactions including the interband and intraband transitions, phonon
excitations, plasmon excitations~\cite{Wooten72}. The energy loss function, $%
L(\omega )$, of ScPtSb (YPtSb) compounds is calculated and shown in 
figure~\ref{fig-13}~(c). This function has the main peak which takes place in
plasma frequency found to be at $21.14$~eV ($16.83$~eV) at zero pressure and $22.92$~eV ($18.21$~eV) at $15$~GPa pressure for ScXSb (YPtSb), respectively, which
corresponds to the abrupt reduction of reflectivity, and the plasma frequency
increases going from ScPtSb to YPtSb.

\section{Conclusions}

In summary, we report  a comprehensive theoretical investigation of the structural, elastic, electronic and optical properties of the half-Heusler compounds ScPtSb and YPtSb using the pseudopotential plane wave in the framework of density functional theory with the GGA-PBEsol functional. (i)~The calculated equilibrium lattice parameters are in good agreement with the available experimental and theoretical data. (ii)~Calculated band structures reveal that ScPtSb is an indirect band gap of $\Gamma-X $ type, while YPtSb is a direct band gap of $\Gamma -\Gamma $ type.  
(iii) The band gaps decrease with an increasing pressure for both studied compounds. (iv)~The nature of the electronic states in the valence and conduction bands was identified through the calculation the diagrams of the TDOS and PDOS. (v)~The elastic constants of both single-crystal and polycrystalline phases of the half-Heusler XPtSb (X=Sc,~Y) were estimated at zero-pressure and under pressure effect up to 15 GPa. It is found that the elastic constants satisfy the mechanical stability criteria in all the considered pressure range. (vi) The Young's modulus, shear modulus, Poisson's ratio, anisotropy
factor, sound velocities, and Debye temperature were estimated and they are in
acceptable agreement with the available data. It is also found that ScPtSb
and YPtSb exhibit a noticeable elastic anisotropy. (vii) According to  the empirical rule
of Pugh, ScPtSb and YPtSb  are ductile materials. (viii) The calculated optical properties, namely the dielectric function, absorption coefficient spectrum, refractive index, extinction coefficient, reflectivity and
energy-loss function were obtained for both compounds. It is found that the static dielectric function at zero pressure increases when Sc atom in ScPtSb is replaced by Y atom. The
dielectric functions increase with an increasing pressure due to the decrease of the optical conductivity and the increase of the band gap. Thus, around zero frequency the optical conductivity of ScPtSb is smaller than that of YPtSb.

\section{Acknowledgements}
This work is supported by PTEA Laboratory, University of Yahia Fares
Medea.


\ukrainianpart

\title[The half-Heusler compounds]%
{Структурні, пружні, електронні та оптичні властивості напів-Гейслерівських сполук  ScPtSb і YPtSb при наявності зовнішнього тиску} %

\author[М. Раджаі]{М. Раджаі\orcid{0000-0002-0313-7155}\refaddr{label1},
	A. Бугемаду\refaddr{label2}, Д. Mауш \refaddr{label2}}
\addresses{
	\addr{label1} Лабораторія фізики експериментальних технологій та їх застосувань, Університет Медеї, Алжир.
	\addr{label2} Лабораторія розробки та дослідження нових матеріалів, Університет Ферхат Аббас Сетиф 1, Сетиф 19000, Алжир.
}

\makeukrtitle

\begin{abstract}
\tolerance=3000%
З метою  дослідження структурних, пружних, електронних та оптичних властивостей напів-Гейслерівських сполук ScPtSb і YPtSb при наявності зовнішнього тиску, що мають кубічну структуру типу
MgAgAs, проведено першопринципні розрахунки з використанням псевдопотенціалу в базисі плоских хвиль у рамках методу апроксимації узагальнених градієнтів. Обчислення проведено з урахуванням спін-орбітальної взаємодії. Розраховані рівноважні сталі ґратки добре узгоджуються з наявними експериментальними і теоретичними значеннями. Жорсткість та механічна стійкість кристалів обговорюється на основі розрахованих  пружних сталих та пов'язаних з ними характеристик, таких як модуль об'ємної пружності, модуль зсуву, температура Дебая, коефіцієнт Пуассона, модуль Юнга та ізотропні швидкості звуку. З розрахованих електронних зонних структур випливає, що в ScPtSb є непряма заборонена зона $\Gamma-X $ типу, в той час як у YPtSb присутня пряма заборонена зона типу  $\Gamma -\Gamma $. Для обидвох сполук ScPtSb та YPtSb досліджено вплив тиску на оптичні властивості, а саме на діелектричну функцію, спектр поглинання, показник заломлення, коефіцієнт екстинкції, коефіцієнт відбиття та спектр енергетичних втрат.

\keywords ScPtSb, YPtSb, метод PP-PW, оптичні властивості, електронні властивості, модулі пружності, першопринципні розрахунки

\end{abstract}

\end{document}